\documentclass[preprint,preprintnumbers,amsmath,amssymb,superscriptaddress,prb]{revtex4}

\usepackage{graphicx}
\usepackage{dcolumn}
\usepackage{bm}
\usepackage{multirow}
\usepackage{color}

\begin{document}


\title{Comment on ``Optical detection of transverse spin-Seebeck effect in permalloy film using Sagnac interferometer microscopy''
}

\author{Johannes Kimling}
\affiliation{Institute of Physics, University of Augsburg, Universit\"atsstrasse 1, 86159 Augsburg, Germany}

\author{Timo Kuschel}
\email{tkuschel@physik.uni-bielefeld.de}
\affiliation{Center for Spinelectronic Materials and Devices, Department of Physics, Bielefeld University, Universit\"atsstrasse 25, 33615 Bielefeld, Germany}

\date{\today}
\maketitle

Measuring spin currents and spin accumulations is challenging. This is mainly because of their small size and the lack of a standard spin detector comparable to conventional ampere meters or voltage detectors. It is all the more pleasing to hear about new methods for detecting spin currents and spin accumulations. A recent example is the paper of McLaughlin, Sun, Zhang, Groesbeck, and Vardeny, who report novel measurements of the transverse spin-Seebeck effect (TSSE) in metals by optical means using a Sagnac interferometer microscope.\cite{McLaughlin2017} However, because no standard procedures or reference measurements are available for verification, critical discussions are crucial for evaluation of such novel experimental approaches.

\par In our comment, we estimate the size of the effect observed by McLaughlin~\textit{et~al.} To illustrate that the estimated size is exceptionally large, we use a comparison with another spin-caloritronic effect: the spin-dependent Seebeck effect. Such a large TSSE is in contradiction with previous experiments that could not resolve a TSSE signal. Therefore, we suggest that signal measured by McLaughlin~\textit{et~al.}~is not or not only caused by the TSSE.

\par \par As stated by McLaughlin~\textit{et~al.}, the TSSE in a ferromagnetic metal thin film is the spontaneous generation of a macroscopic spatial distribution of accumulation of spin-polarized carriers by an in-plane temperature gradient in the presence of a magnetic field.\cite{McLaughlin2017} This effect was first reported by Uchida~\textit{et~al.}~, who explained the observed macroscopic length-scale of the TSSE with a simple mechanism based on conduction electrons that have a spin-dependent Seebeck coefficient in ferromagnetic metals.\cite{Uchida2008} This explanation was soon refuted, e.g., by Hatami~\textit{et~al.}~and by Scharf~\textit{et~al.}, who clarified that the thermally-driven spin accumulation cannot persist over distances longer than the spin-diffusion length, which in ferromagnetic metals is restricted to nanometer length-scales.\cite{Hatami2010,Scharf2012} Because of the enormous influence of the original TSSE report in the growing field of spin caloritronics,\cite{Bauer2012} the nanoscale effect was renamed into spin-\textit{dependent} Seebeck effect and new theories based on collective electron effects were developed to explain the observed macroscopic length-scale of the TSSE.\cite{Xiao2010,Adachi2013,Rezende2014}   

\par To explain their observations, McLaughlin~\textit{et~al.}~refer to a phonon-magnon drag mechanism described in the new TSSE theories. However, the TSSE theories of Xiao~\textit{et~al.}~and of Adachi~\textit{et~al.}~describe a different experiment that includes F/N interfaces.\cite{Xiao2010,Adachi2013} Furthermore, Xiao~\textit{et~al.}~state that their TSSE theory fails for Py,\cite{Xiao2010} and the theory of Rezende~\textit{et~al.}~describes the longitudinal SSE, not the transverse SSE. Hence, it is unclear if the new TSSE theories can explain the experiment of McLaughlin~\textit{et~al.} Below, we argue that the size of the effect observed by McLaughlin~\textit{et~al.}~makes it even more challenging to explain this effect with the new TSSE theories. 

\par McLaughlin~\textit{et~al.}~state to measure changes in Kerr rotation that originate from two effects: (i) change of saturation magnetization caused by a temperature increase ($\Delta M_{T\rm }$); (ii) generation of spin accumulation or nonequilibrium magnetization caused by the TSSE ($\Delta M_{\rm TSSE}$). However, McLaughlin~\textit{et~al.}~do not discuss the size of the TSSE they observe, which would require knowledge of the coefficient that relates the change in Kerr rotation to the change in nonequilibrium magnetization ($\Delta\theta_{\rm K}/\Delta M_{\rm TSSE}$). In the following, we provide a rough estimation of the nonequilibrium magnetization that McLaughlin~\textit{et~al.}~state to observe. 

\par According to Fig.~3 of their paper, application of a temperature difference of 1~K between the ends of a millimeter sized sample produces absolute changes in Kerr rotation of approximately 500~nrad caused by the TSSE (effect (ii) above). According to Table~I in their paper, the absolute change in Kerr rotation produced by a temperature change of the entire sample of 1~K is also approximately 500~nrad (effect (i) above). Let's assume that both effects (i) and (ii) are measured with the same sensitivity, i.e., $\Delta\theta_{\rm K}/\Delta M_{\rm TSSE}\approx\Delta\theta_{\rm K}/\Delta M_{\rm T}$. Then, we can estimate the nonequilibrium magnetization with the change of the saturation magnetization when changing the temperature of the entire sample by 1~K.

\par For bulk Py at room temperature, the absolute change of saturation magnetization per 1~K temperature change is approximately 300~A~m$^{-1}$.\cite{Mauri1989} Hence, based on the above reasoning, McLaughlin~\textit{et~al.}~observe a nonequilibrium magnetization of the order of 300~A~m$^{-1}$ caused by the TSSE. Note that this value is a lower limit, because change in Kerr rotation due to spin accumulation is much smaller than change in Kerr rotation due to change in magnetization, i.e., $\Delta\theta_{\rm K}/\Delta M_{\rm TSSE}\ll\Delta\theta_{\rm K}/\Delta M_{\rm T}$,\cite{Choi2014C} and because the temperature coefficient of the magnetization of a Py thin film is larger compared to bulk Py.\cite{Mauri1989}

\par To better understand the size of the estimated nonequilibrium magnetization, we use the spin-dependent Seebeck effect in a 50-nm-thick Py layer with an out-of-plane temperature gradient for comparison. Note that we do not propose the spin-dependent Seebeck effect as an alternative explanation, but to illustrate that the effect observed by McLaughlin~\textit{et~al.}~is exceptionally large. Using Eq.~(49) from Ref.~\onlinecite{Scharf2012}, the spin-dependent Seebeck effect generates a nonequilibrium magnetization of\cite{Scharf2012} 
\begin{equation}
\Delta M = \frac{S_{\rm S}}{2}\lambda_{\rm S}\frac{\Delta T}{h}\frac{\sinh(x/\lambda_{\rm s})}{\cosh(h/2\lambda_{\rm S})}\frac{N\mu_{\rm B}e}{2},
\end{equation}
where $S_{\rm S}$, $\lambda_{\rm S}$, $N$, and $h$ are spin-dependent Seebeck coefficient, spin-diffusion length, electron density of states, and thickness of the Py layer. The thickness of the Py layer is $h=50$~nm; we estimate $N = 3\gamma/(\pi^2k_{\rm B}^2)\approx1.7\times10^{48}$~J$^{-1}$~m$^{-3}$ using $\gamma = 1064$~J~m$^{-3}$~K$^{-2}$ of Ni.\cite{Kittel} Assuming a spin-dependent Seebeck coefficient of $S_{\rm S} \propto 4.5\times10^{-6}$~V~K$^{-1}$ and a spin-diffusion length of $\lambda_{\rm S}\sim 5$~nm,\cite{Dejene2012} the temperature difference across the 50~nm thick Py layer required to produce a nonequilibrium magnetization of $\Delta M = 300$~A~m$^{-1}$ at the surface of the Py layer is
\begin{equation}
\Delta T = \frac{4h}{S_{\rm S}\lambda_{\rm S}N\mu_{\rm B}e}\frac{\cosh(h/2\lambda_{\rm S})}{\sinh(x/\lambda_{\rm s})}\Delta M\approx\frac{4h}{S_{\rm S}\lambda_{\rm S}N\mu_{\rm B}e}\Delta M\approx1058\rm~K
\end{equation}
Considering the small temperature gradients of the order of 1~K over a millimeter distance involved in the SSE experiment of McLaughlin~\textit{et~al.}, this example illustrates that the TSSE must be exceptionally large to explain the conjectured nonequilibrium magnetization of $\Delta M = 300$~A~m$^{-1}$. Such a large TSSE would contradict prior experiments with Pt/Py bilayers, because the result of the prior experiments was that after control of measurement artifacts, in particular out-of-plane temperature gradients, possible TSSE signals were not observed, i.e., were either absent or below the measurement resolution.\cite{Huang2011,Avery2012,Schmid2013,Meier2013,Shestakov2015}

\par Before we conclude our comment, we briefly address the incorrect statement that our previous paper (Ref. 36 in the paper of McLaughlin~\textit{et~al.}) would report time-resolved magneto-optic Kerr effect measurements of the spin-dependent Seebeck effect.\cite{McLaughlin2017} We clarify that our previous paper reports optical measurements of the longitudinal SSE.\cite{Kimling2017b}  

\par To conclude, we estimate that McLaughlin~\textit{et~al.}~observe a nonequilibrium magnetization of the order of $\Delta M = 300$~A~m$^{-1}$ at the surface of the Py layer. To produce a comparable amount of spin accumulation in the 50-nm-thin Py layer only from the spin-dependent Seebeck effect would require a temperature difference of the order of 1000~K over the thickness of the Py layer. Considering the null results from previous TSSE experiments, we suggest that the signal measured by McLaughlin~\textit{et~al.}~is not or not only caused by the TSSE.


\end{document}